\documentclass[prl,aps,twocolumn,showpacs,groupedaddress,superscriptaddress]{revtex4}

\usepackage{epsfig}

\def\mymarginpar#1{\marginpar{\tiny\it\raggedright #1}}
\begin{document}

\def\mymarginpar#1{\marginpar{\tiny\it\raggedright #1}}

\newcommand{\be}{\begin{equation}}
\newcommand{\ee}{\end{equation}}
\newcommand{\beq}{\begin{eqnarray}}
\newcommand{\eeq}{\end{eqnarray}}

\newcommand{\w}{\omega}
\newcommand{\W}{\Omega}
\newcommand{\g}{\gamma}
\newcommand{\G}{\Gamma}
\newcommand{\E}{\hat{\cal E}}
\renewcommand{\a}{\hat a}
\newcommand{\s}{\sigma}
\newcommand{\n}{\bar{n}}
\newcommand{\ket}{\rangle}
\newcommand{\bra}{\langle}
\newcommand{\nn}{\nonumber}
\providecommand{\kb}[2]{\lvert\,#1\rangle\langle#2\rvert}
\newcommand{\me}{\mathrm{e}}
\newcommand{\dif}{\mathrm{d}}
\newcommand{\bnn}{\begin{eqnarray*}}
\newcommand{\enn}{\end{eqnarray*}}

\newcommand{\ds}{\displaystyle}
\newcommand{\dd}{\partial}
\newcommand{\dt}{\ds\frac{\dd}{\dd t}}
\newcommand{\dz}{\ds\frac{\dd}{\dd z}}
\newcommand{\D}{\ds\left(\frac{\dd}{\dd t} + c \frac{\dd}{\dd z}\right)}
\newcommand{\nt}[1]{{\it #1}}
\newcommand{\Rb}{{\it Rb}}
\newcommand{\Fig}{Fig.\ }

\newlength{\textwidthm}
\setlength{\textwidthm}{\columnwidth}
\addtolength{\textwidthm}{-\parindent}
\addtolength{\textwidthm}{-\parindent}

\newlength{\texthwidthremark}
\setlength{\texthwidthremark}{\columnwidth}

\newcommand{\remark}[1]{\\ \hspace*{-.5cm}
\parbox[t]{\texthwidthremark}{
{\large\it #1\\}}
}

\title{Observation of narrow Autler-Townes components in the resonant response of a dense atomic gas}

\author{Vladimir A. Sautenkov}
\affiliation{
    Department of Physics and Institute of Quantum Studies,
    Texas A\&M University,
    College Station, Texas 77843-4242
}
\affiliation{
    P.  N.  Lebedev Institute of Physics, 119991 Moscow, Russia
}

\author{Yuri V. Rostovtsev}
\affiliation{
    Department of Physics and Institute of Quantum Studies, Texas A\&M University,
    College Station, Texas 77843-4242
}

\author{Eric R. Eliel}
\affiliation{ Huygens Laboratory, Leiden University, P.O.
Box 9504, 2300 RA Leiden, The Netherlands }

\date{\today}

\begin{abstract}
We have experimentally studied the reflection of a weak
probe beam from a dense atomic potassium vapor in the
presence of a strong laser field tuned to the atomic
resonance transition. We have observed an Autler-Townes
doublet under hitherto unexplored conditions, namely that
the Rabi frequency induced by the strong laser
field is much smaller than the self-broadened width
 of the resonance transition of the unexcited
vapor. We attribute our observation to a reduction of the
atomic decoherence by the strong drive field. We present a
theoretical model of nonlinear processes in a dense atomic
gas to explain the observed results.
\end{abstract}


\maketitle 

Spectral line broadening is a universal
phenomenon and a multitude of techniques has been devised
to reduce or eliminate it altogether, enabling the study of
spectral features that would, otherwise, remain
hidden~\cite{Meschede,Levenson,boyd,Svanberg}. In particular
this applies to atomic or molecular gases where the
collision rate can be made sufficiently small by, for
instance, rarefying the vapor. However, when the interest
lies with the (strongly) interacting gas, line broadening becomes the
essence, and the study of the  width
of the fundamental resonance transitions in high-density
atomic vapors is well-documented~\cite{Lewis,Leegwater}.

\begin{figure}[ht]
    \includegraphics[
        angle=0,
        width=0.95\columnwidth
    ]
    {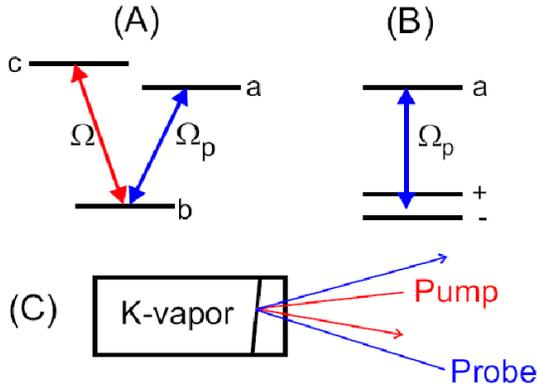}
        \caption{
(Color online) Energy-level scheme for potassium. The bare
atomic states are shown in (A), where level $a$ corresponds
to the $4p\ ^2P_{1/2}$ excited state, level $b$ to the $4s\
^2S_{1/2}$ ground state, and level $c$ to the  $4p\
^2P_{3/2}$ excited state. The strong drive field, with Rabi
frequency $\Omega$, couples states $b$ and $c$, while the
probe laser beam, with Rabi frequency $\Omega_{\mathrm{p}}$
is scanned over the $b-a$ transition. Frame (B) shows the
probe transition in the dressed basis, the states being
written as $|+\rangle$ and $|-\rangle$, and (C) provides a
sketch of the experimental setup. \label{setup}
    }
\end{figure}

Although, in the general case, the broadening of the resonance line has many 
contributions \cite{line-broadening}, 
for dense homogeneous gases, the major contribution is caused by the
resonant dipole-dipole interaction between ground- and
excited-state atoms of the same species, and is known as
self- or resonance broadening~\cite{Lewis,Leegwater}. 
The interaction is long
range, so that already at reasonably modest densities of
order $10^{17}\ \mathrm{cm}^{-3}$ one leaves the
binary-collision regime and multi-perturber effects may
come into play~\cite{Leegwater, West}. 
It has, for instance, been shown that the
Zeeman effect is modified at atomic densities of this
order~\cite{Kampen1}. We note that multi-perturber effects were
observed in high presure buffer gas~\cite{Ref1, Ref2, Ref3} when the duration
of collisions cannot be neglected.

In a high-density vapor the atom responds not just to the
externally applied EM field but also to the field
re-radiated by other atoms in the vapor, i.e. it responds
to the \textit{local} field. As first shown by Lorentz,
this causes the position of the resonance to be
shifted~\cite{Lorentz}. During the last decade this
local-field shift, named after Lorentz, has been studied by
both frequency~\cite{Boyd1,Boyd2,Cooper,Stroud,Sautenkov,Kampen2}
and time-domain~\cite{Cundiff} techniques. It is
proportional to the atomic density, and of the same order
of magnitude as the line width~\cite{Friedberg}.

It has been predicted that the Lorentz local-field shift
depends on the degree of (incoherent) excitation of the
vapor~\cite{Friedberg,Manassah}, a prediction that
stimulated the study of a variety of nonlinear optical
phenomena in dense atomic vapors, such as piezophotonic
switching and lasing without inversion~\cite{Manka}. The
experimental demonstration of the excitation dependence of
the Lorentz shift~\cite{Sautenkov,Kampen2} brought to light
that the self-broadened width itself is also excitation
dependent, an effect that could be explained in terms of a
quasistatic (multi-perterber, quasi-molecular)
collision model~\cite{Leegwater,Sautenkov}.
Recently, it was shown by using a time-domain technique, that non-Markovian collision dynamics can be observed in these high density resonantly-broadened vapors, indicating that the duration of a collision can no longer be
neglected~\cite{cundiff05}.

\begin{figure}[ht]
    \includegraphics[
        angle=0,
        width=0.95\columnwidth
    ]
    {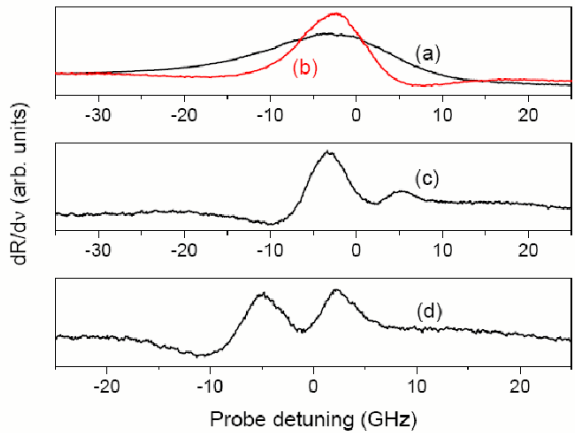}
    \caption{(Color online)
Frequency-modulated reflectivity spectra of a potassium
vapor at a density of $ 4.9\times 10^{17}$ cm$^{-3}$, as
measured on the D$_1$ transition. Curves(a) and (b) display
the FM reflection spectrum of the vapor when the pump laser
is switched off, and when it operates at a power of
$\approx 0.5$ W and is tuned far (130 GHz) above the D$_2$
transition. When the detuning of the pump laser is reduced
to $\approx 3$ GHz the spectral line in the FM spectrum
becomes an Autler-Townes doublet (curve (c)).
 Curve (d) shows the Autler-Townes
components at zero pump detuning. \label{fig2}
    }
\end{figure}

The experimental study of resonance broadening remains a
topical issue in view of theoretical claims that, because
of the long-range nature of the interaction, many-body
effects play a role at all densities~\cite{Leegwater}. 
Here we show that at a higher number
density of $4.9\times 10^{17}$ cm$^{-3}$ the atomic
ensemble can be made to evolve coherently as if it is a
two-level system with intensity dependent line width. 
We deduce this behavior from our
measurements of the Autler-Townes splitting~\cite{Autler} of the
potassium atomic ground state. 
Observation of the Autler-Townes splitting of optical transition in delute 
gases were previously reported in \cite{toschek}. 
Essential to our observation
is the effect of collision-induced line
narrowing~\cite{Stroud,Sautenkov,Kampen2}.

At the number density of our experiment, the self-broadened
line width of both resonance lines of potassium is
considerably larger than either their hyperfine structure
or their Doppler width. Under these conditions the level
structure of potassium is well-represented by a three-level
scheme, shown in Fig.~\ref{setup}. A strong driving field
(Rabi frequency $\Omega$) couples levels $b$ and $c$, while
a weak probe field, with Rabi frequency
$\Omega_{\mathrm{p}}$ is scanned over the $a\leftrightarrow
b$ transition.

The experiment is performed in reflection (see
Fig.~\ref{setup}C), using an experimental arrangement
similar to that of Ref. \cite{Kampen2}. The strong coupling
field is produced by a CW Ti-sapphire laser tuned close to
the $D_2$-line ($\lambda= 766$ nm) of atomic potassium. The
laser beam, having a maximum power of $\approx 0.5$ W, has its
focus (the area of the beam is $5\ 10^{-4}$ cm$^2$) 
at the window of a vapor cell that
is constructed entirely out of sapphire. A weak probe beam
from an external cavity diode laser is focussed to overlap
with the pump beam at the entrance window of the cell.  The
probe laser is scanned over the $D_1$-line at $\lambda=
770$ nm, at a power level of $\approx 10^{-4}$ W, far below
the onset of saturation
(saturation parameter~\cite{boyd} is $< 0.1$). 
Both the pump and probe beams are
linearly polarized and are incident on the sapphire cell at
near-normal incidence. While being scanned, the probe laser
is frequency modulated over a range of 100 MHz at a
frequency of 400 Hz. The reflected probe beam is captured
by a photodiode whose signal output is fed to a lock-in
amplifier.  Our experiment runs at an atomic density of
$4.9\times 10^{17}$ cm$^{-3}$, yielding  a self-broadened
linewidth $\G_{\mathrm{self}}/2\pi$ of 28.4  GHz for the
D$_1$ line~\cite{Kampen1}.

Figure~\ref{fig2} shows the measured FM-modulated spectra.
The zero on the frequency axis refers to the center
frequency of the absorption spectrum of a low density ($N \simeq
10^{11}$ cm$^{-3}$) potassium vapor in a reference glass
cell. Curve (a) displays the reflection spectrum when the
pump beam is switched off, i.e. it shows the FM linear 
reflectivity of the sapphire/high-density vapor
interface. When the pump laser is switched on, operates at
an output level of $0.5$ W, and is tuned 130 GHz
above the D$_2$ line, the FM spectrum of curve (b) is
obtained. At this large detuning the pump-laser beam
penetrates a considerable distance into the vapor cell. Due
to radiation trapping the decay of the excited-state
population is slowed down and, as a result of radiative
transport, the vapor near the sapphire window becomes
homogeneously and incoherently excited. In line with
earlier observations (\cite{Kampen2}) one notices that the
FM spectrum of the strongly excited vapor (curve (b)) is
narrower than its unexcited counterpart (curve (a)), and
that their centers do not coincide. However, both spectra
are described by the same spectral function, with adjusted
values for the Lorentz local-field shift and self-broadened
width~\cite{Stroud,Sautenkov,Kampen2}.

The spectral width $\G_\mathrm{self}$ of the resonance line
can be estimated from the interval $\Delta\w_\mathrm{mm}$
between the maximum and minimum of the reflection spectrum
or, equivalently,  from the zeros of the FM spectrum.
 For the $D_1$ transition one has
$\G_\mathrm{self} =
0.87\Delta\w_\mathrm{mm}$~\cite{Kampen1, Kampen2}. By this
method, we determine the line width for the unexcited vapor
as $31$ GHz (Fig. \ref{fig2}a), in  reasonable agreement with the value
$\G_\mathrm{self}/2\pi = 28.4$ GHz, calculated using the
known self-broadening coefficient and our estimate for the
atomic density. Similarly, we estimate the line width for
the incoherently excited  vapor as just $11$ GHz (Fig. \ref{fig2}b). 
Hereby we
show that the excitation dependence of the self-broadened
line width is a valid concept also at the vapor density of
the present experiment.

In order to observe coherent effects we tune the pump laser
close to the center of the atomic line. Curves (c) and (d)
show the experimental FM reflectivity spectra for a pump
detuning of 3 and 0 GHz, respectively. In both cases one
observes that the single resonance of curve (b) is split
into an Autler-Townes doublet~\cite{Autler,toschek}. The asymmetry
of the doublet in curve (c) reflects the fact that, for
this curve, the pump laser is tuned slightly above the
D$_2$ resonance transition, while the near-perfect symmetry
of curve (d) shows that the pump laser is, in that case,
tuned on resonance. Note that, as a result of the various
line shifts, there is no easy method, beyond looking at the
symmetry of the Autler-Townes doublet, to establish whether
the pump laser is tuned exactly to resonance or not.

To obtain a good estimate for the width and separation of
the the components of the Autler-Townes doublet we have
made a non-linear fit of the experimental data of curve (d)
in Fig.~\ref{fig2}, yielding a width of 8.5 GHz and
 a splitting of 7.6 GHz.

It is well known that the separation between the two
components of the Autler-Townes doublet equals the Rabi
frequency induced by the pump laser~\cite{boyd,Cohen}.
For the conditions of the present exeriment where hyperfine
and Zeeman substructure can be ignored, the Rabi frequency
$\W$ for the potassium $D_2$ line can be expressed as
\begin{equation}
 \W/2\pi =  8\times 10^7
\sqrt{I\left[{\mathrm{W}\over\mathrm{cm}^2}\right]}\;\; 
\mathrm{Hz},\label{eq2} \end{equation} where
$I= {cE^2/(8\pi)}$ is the pump laser intensity 
(in units of W/cm$^2$). Here the
Rabi frequency is written as \be \W = E\wp_x/\hbar,
\label{eq3}\ee with $E$ the optical electric field and
$\wp_x$ the projection of the atomic dipole moment
$\vec{\wp}$ along that field. The value of $\wp_x=\wp/3$
has been obtained from experimental data for the
spontaneous lifetime \cite{Sobelman}. Using Eq.~(\ref{eq2})
we calculate the Autler-Townes splitting to be 8 GHz at a
pump power of 0.5 W, in satisfying agreement with the
experimental value (7.6 GHz).

\begin{figure}[ht]
    \includegraphics[
        angle=0,
        width=0.95\columnwidth
    ]
    {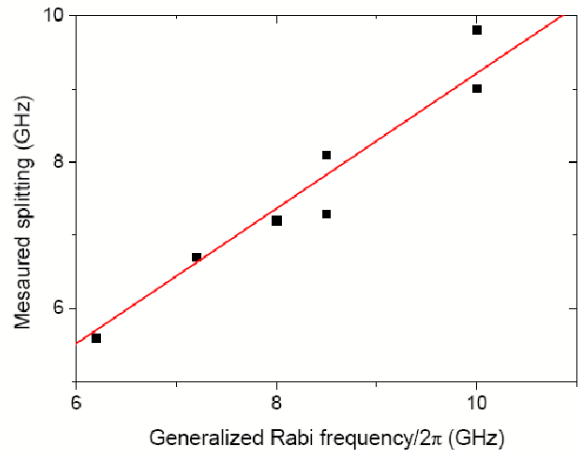}
        \caption{The measured Autler-Townes splitting versus generalized Rabi frequency.
    The solid line shows the result of a linear fit.
\label{fig3}
    }
\end{figure}

We have recorded the Autler-Townes doublet for several
values of the detuning $\Delta$ and intensity of the pump
laser. When $|\Delta|$ increases  one of the components of
the doublet becomes stronger while the other fades away.
When $|\Delta/2\pi|>6$ GHz the weak component can no longer
be measured, and only a single resonance is observed. For
this case, where the pump laser is detuned from exact
resonance, the Autler-Townes splitting is given by the
generalized Rabi frequency~\cite{Cohen}
 \be \tilde{\W} = \sqrt{\W^2 +
\Delta^2}. \label{eq4}\ee The measured values of the
Autler-Townes splitting  as a function of the generalized
Rabi frequency $\tilde{\Omega}$  are presented in
Fig.~\ref{fig3}. The three points for $\tilde{\Omega}/2\pi
\le 8$ GHz have been measured at zero detuning  and  pump
laser powers equal to 0.3, 0.4 and 0.5 W, respectively.
The points at $\tilde{\Omega}/2\pi\ge 8.5$ GHz were
measured by using a pump power equal to 0.5 W and
detunings of  $\Delta/2\pi = \pm 3$ GHz and $\pm 6$ GHz,
respectively. Also shown is a linear fit through the data;
its slope equals 0.92, in quite good agreement with the
expected value, equal to 1 (the error of slope is 2\% and 
the difference between theoretical and experimental curves is
about 6\% due to systematic uncertanties).

\begin{figure}[hb]
    \includegraphics[
        angle=0,
        width=0.95\columnwidth
    ]
    {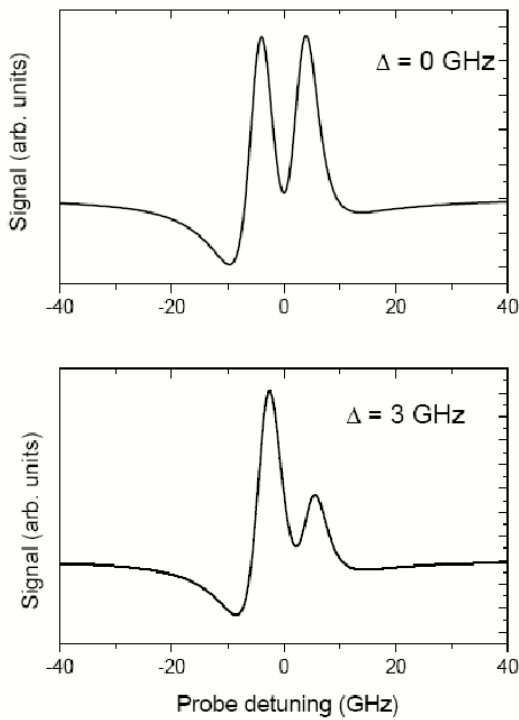}
      \caption{
Results of the theoretical model for the FM-reflection
spectrum of a vapor of coherently driven V-type three-level
atoms at high density, using Eqns.~(\ref{eq6}) and
(\ref{eq8}). \label{theory}
    }
\end{figure}

To explain our results we use the optical Bloch equations
for a V-type three-level atom, suitably modified for the
fact that the system under study is a dense rather than a
dilute vapor. From these equations we evaluate the
susceptibility of the probe at frequency $\w_p$
as~\cite{sz}
 \be \chi(\w_p) = i\frac{3\lambda^2 \g}{8\pi}{(n_{a}-n_b) +
\ds{\W^2\over\G_{cb}\G_{ca}} (n_c-n_b) \over\G_{ab} +
\ds{\W^2\over\G_{ca}}}. \label{eq6}\ee Here the level
populations $n_{a}$, $n_b$ and $n_{c}$ depend on the Rabi
frequency of the driving field $\W$, $\g$ is the radiative
rate on the $a\rightarrow b$ transition, and the
generalized coherence decay rates are $\G_{cb} =
\tilde\g_{cb} + i\delta_{cb}$, $\G_{ca} = \tilde\g_{ca} +
i\delta_{ca}$, $\G_{ab} = \tilde\g_{ab} + i\delta$. The
frequency of the pump is denoted as $\bar\w$, and its
detuning is given by $\delta_{cb} = \w_{cb} - \bar\w +
\Delta\w_L + \Delta\w_c$, with $\Delta\w_L$ the Lorentz
local field shift, and $\Delta\w_c$ the collisional
frequency shift, which is independent of the level of
excitation \cite{Kampen2, Boyd1,Boyd2}. The detuning of the
probe from resonance is $\delta = \w_{ab} - \w_p +
\Delta\w_L + \Delta\w_c$, with $\w_p$ the frequency of the
probe laser. For the two photon detuning we write
$\delta_{ca} = \w_{ca} - \bar\w + \w_p + \Delta\w_L +
\Delta\w_c$.  For $\tilde\g_{cb}$, $\tilde\g_{ca}$ and
$\tilde\g_{ab}$ we write $\tilde\g_{ab} = \g/2 +
\G_\mathrm{self} \simeq \G_\mathrm{self}$, $\tilde\g_{cb} =
\g/2 + \G_\mathrm{self} \simeq \G_\mathrm{self}$, and
$\tilde\g_{ca} = \g + \G_\mathrm{self} \simeq
\G_\mathrm{self}$,
   since the
self-broadened line width is much larger than the natural
width of the transitions.

For the self-broadened width we write~\cite{Sautenkov,
Kampen2} \be \G_\mathrm{self} = kn_{b}, \ee with $k$ the
self-broadening coefficient~\cite{Lewis}. The Lorentz
local-field shift is given by  \cite{Friedberg, Manassah}
\be \Delta\w_L =  \frac{k}{3}(n_b-n_a). \ee

The reflectivity of the coherently driven dense vapor can
now be evaluated using  \be R(\w) = \left|{n(\w) - n_0\over
n(\w) + n_0}\right|^2, \label{eq8} \ee with
$n(\w)=\sqrt{1+4\pi\chi(\w)}$ and $n_0$ the refractive indices of the
vapor and  of the window material
of the vapor cell, respectively. Using Eq.~(\ref{eq8}) and
appropriate values for the experimental parameters we calculate the FM
reflection spectra, as shown in Fig.~\ref{theory}, for a
pump detuning of zero and 3 GHz, respectively, and a pump Rabi frequency of 12 GHz. Fully in
line with the experimental result our theoretical model
predicts a well-resolved Autler-Townes doublet under
circumstances that the Rabi frequency is much smaller than
the unsaturated self-broadened line width.

It is worth to mention here that, even though the experiments and theoretical 
analysis have been done for conditions corresponding to hot dense gases, 
it will be interesting to study discussed multy-perturber effects in dense 
cold atomic and molecular gases, and possible influence on coherent 
backscattering in cold atoms~\cite{cold-atoms-ab}.

In conclusion, we have observed the Autler-Townes splitting
of the atomic ground state in the reflection spectrum of a dense potassium vapor in a regime where the separation of the two Autler-Townes components, as given by the Rabi frequency, is much smaller than the nominal  self-broadened width of the transition. We attribute our observation to the fact that the strong drive field, in addition to inducing coherent evolution of the atomic levels, also causes a dramatic
narrowing of the spectral line. Taking into account the
excitation dependence of both the local-field shift and the resonance line
width we obtain good
agreement between our experimental results and an effective three-level model based on modified Bloch equations.

We wish to acknowledge S.T. Cundiff, 
H. van Kampen, S. Mukamel, and J. P. Woerdman for
useful discussions and beneficial help. This work was done, in part, 
while V.A.S. was a visitor at Leiden University as part of the research
program of the "Stichting voor Fundamental Onderzoek der
Materie". V.A.S. and Yu.V.R. also gratefully acknowledge
the support from the NSF, the Office of Naval Research, the
Air Force Research Laboratory (Rome, NY), Defense Advanced
Research Projects Agency, and the Robert A.\ Welch
Foundation (Grant \#A-1261).



\end{document}